# Deep Multi-path Network Integrating Incomplete Biomarker and Chest CT Data for Evaluating Lung Cancer Risk

Riqiang Gao[a*], Yucheng Tang[a], Kaiwen Xu[a], Michael N. Kammer[b], Sanja L. Antic[b], Steve Deppen[b], Kim L. Sandler[b], Pierre P. Massion[b], Yuankai Huo[a], Bennett A. Landman[a,b]

[a] Electrical Engineering and Computer Science, Vanderbilt University, Nashville, TN, USA 37235
[b] Vanderbilt University Medical Center, Nashville, TN, USA 37235
(*Corresponding Author: riqiang.gao@vanderbilt.edu)

## ABSTRACT

Clinical data elements (CDEs) (e.g., age, smoking history), blood markers and chest computed tomography (CT) structural features have been regarded as effective means for assessing lung cancer risk. These independent variables can provide complementary information and we hypothesize that combining them will improve the prediction accuracy. In practice, not all patients have all these variables available. In this paper, we propose a new network design, termed as multi-path multi-modal missing network (M3Net), to integrate the multi-modal data (i.e., CDEs, biomarker and CT image) considering missing modality with multiple paths neural network. Each path learns discriminative features of one modality, and different modalities are fused in a second stage for an integrated prediction. The network can be trained end-to-end with both medical image features and CDEs/biomarkers, or make a prediction with single modality. We evaluate M3Net with datasets including three sites from the Consortium for Molecular and Cellular Characterization of Screen-Detected Lesions (MCL) project. Our method is cross validated within a cohort of 1291 subjects (383 subjects with complete CDEs/biomarkers and CT images), and externally validated with a cohort of 99 subjects (99 with complete CDEs/biomarkers and CT images). Both cross-validation and external-validation results show that combining multiple modality significantly improves the predicting performance of single modality. The results suggest that integrating subjects with missing either CDEs/biomarker or CT imaging features can contribute to the discriminatory power of our model ($p < 0.05$, bootstrap two-tailed test). In summary, the proposed M3Net framework provides an effective way to integrate image and non-image data in the context of missing information.

## 1. INTRODUCTION

Lung cancer is the most deadly cancer in United States [1]. Early lung cancer detection with low-dose computed tomography (CT) of the chest can reduce the relative risk of lung cancer mortality by 20% [2]. On the other hand, Clinical data elements (CDEs) and biomarkers have also been widely used for cancer risk estimation in researches and practice [3] [4]. The National Lung Screening Trial (NLST) selected subjects according to CDEs (e.g., age >=55, pack-years >= 30) [5]. The Mayo team introduced model that identifies malignancy of nodules with clinical data and radiological characteristics of pulmonary nodules [3]. Kammer et al. [4] empirically validated the blood marker hs-CYFRA 21-1 for improving the diagnosis of lung cancer. Several methods (e.g., [6]) integrated the CT reader information (e.g., nodule size) and biomarkers to predict the lung cancer risk. With flourishing of deep learning in computer vision fields, the deep convolutional neural network (CNN) has been widely adopted to extract medical image features. Liao et al. proposed a 3-D CNN with pulmonary nodule detection and nodule malignancy classification modules for lung cancer detection [7], which won the Data Science Bowl Kaggle challenge (https://www.kaggle.com/c/data-science-bowl-2017). Several subsequent methods extended Liao's framework to serial CT images [8][9][10] and multi-task networks [11] [12]. As validated, the CDEs/biomarkers and CT images are helpful for lung cancer detection. Yet, not all the subjects have the complete multi-modality data of CDEs, biomarkers and chest CT, which brings the challenge to learn a deep network with missing data. Note that in this paper, CDEs and biomarkers are regarded as one modality and CT image as another. For clarity, the CDEs/biomarkers will be termed as biomarkers in the following.

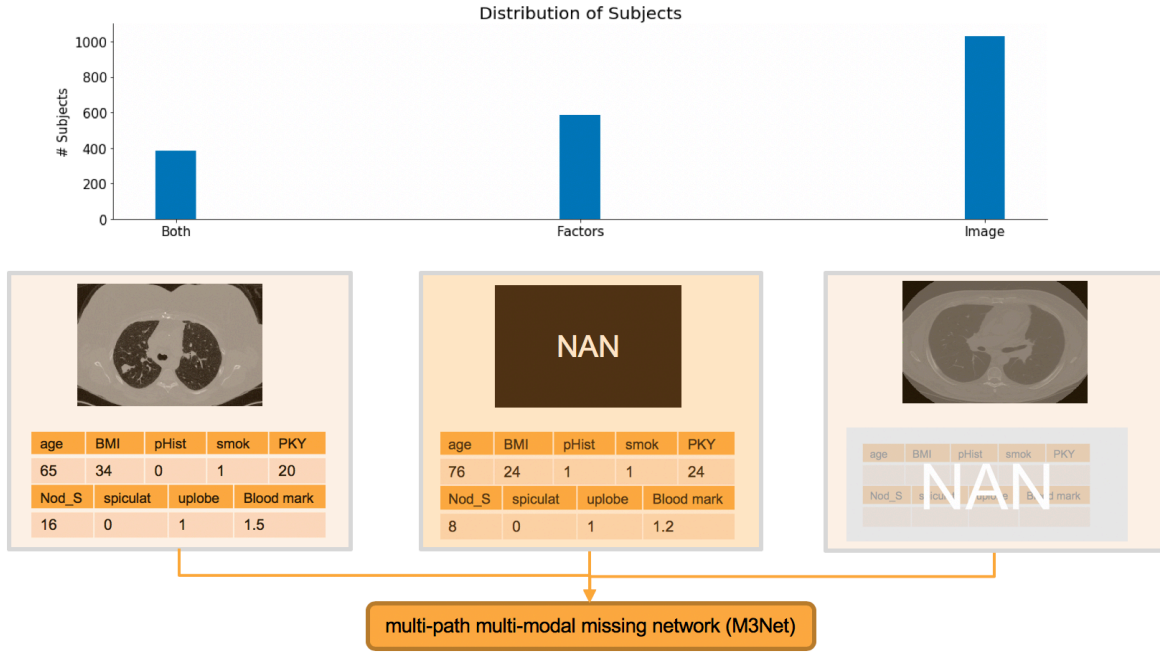

**Fig 1**. The intuition of the proposed M3Net. In practice, not all subjects have both clinical variables, biomarker and CT images. Our network takes the available data, including complete data and data with missing modality, to train a uniform deep network. During the test phase, our model can predict lung cancer risk with incomplete data.

Research of learning with missing data can be dated back to 1970s [13], where the generality and problem of missing data are introduced. The solutions for missing data can be broadly categorized to two main classes: *deletion* and *imputation*. There were several ad-hoc solutions raised for missing data, including listwise deletion, pairwise deletion, mean imputation, last observation carried forward, which were well introduced and organized by [14]. The leading methods of *imputation* can be categorized as either discriminative or generative [15]. The discriminative imputation models include MICE [16], MissForest [17]. The generative imputation models involve VAE variants [18] and GAN variants [15][19]. One drawback of majority existed imputation models is that they are based on the same type of data (e.g., impute images based on observed images). In practice, the missing data are usually heterogenous and may include image and non-image types. The ad-hoc solutions of *listwise deletion* and *pairwise deletion* can be extended to the field of machine learning. A straightforward operation is to delete the samples with missing component when training a machine learning algorithm, which is the *listwise deletion* extension. However, the imperfect data may also contribute, e.g., Yang et al. [12] ignored the backpropagation of one task when its related label is missing when training a multi-label network for chest CT.

In contrast to previous research with missing values [15] [18], image inpainting [19], and image modality missing [20], we focus on the missing modality across image and non-image data in this paper. We propose a new network, termed as multi-path multi-modal missing network (M3Net), to cope with missing data in lung cancer risk estimation, as shown in Figure 1. Generally, the number of subjects with one modality (either biomarkers or CT) is relatively high, while the fewer subjects with both complete modalities of biomarker and CT. The M3Net integrates (1) multi-modality data (i.e., CT images and biomarkers) and (2) data with missing modality in an end-to-end training network with multi-path. The data with missing modality can be included in both training and testing sets. Our datasets come from three sites: Vanderbilt University Medical Center (VUMC), University of Colorado Denver (UCD), the Detection of Early Cancer Among Military Personnel (DECAMP) and University of Pittsburgh Medical Center (UPMC). Our network is evaluated by cross-validation with available subset of VUMC + UCD + DECAMP (termed as VDD in the following), and external validated with the UPMC cohort. Our M3Net shows superior performance over the single modality predictions and learning without incomplete data.

Table 1. The number of subjects in the cohorts

| Cohorts | VDD | UPMC |
|---|---|---|
| Total Subjects | 1232 | 99 |
| With biomarkers | 585 | 99 |
| With images | 1030 | 99 |
| With image and biomarkers | 383 | 99 |

## 2. METHODS

### 2.1 Data Cohorts Description

The data used in our experiments is subset of the project the Consortium for Molecular and Cellular Characterization of Screen-Detected Lesions (MCL). We have collected the subjects from VUMC, UCD, DECAMP and UPMC under Institutional Review Board (IRB) supervision. The subjects from VUMC, UCD and DECAMP are combined as the cohort VDD. The five-fold cross-validation is performed in the VDD cohort, and UPMC cohort is used as external validation.

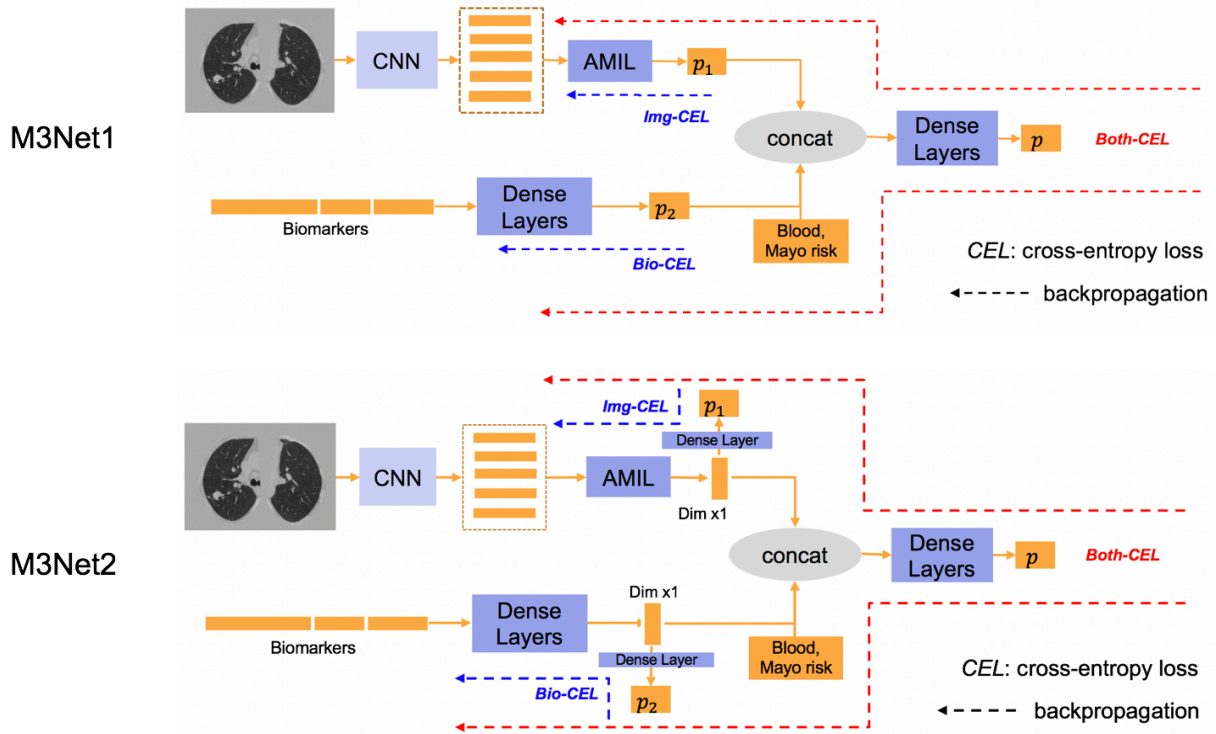

**Fig 2.** The framework of the proposed M3Net, including two versions M3Net1 (upper) and M3Net2 (bottom). The M3Net includes three paths, one for CT image and another for biomarkers, and one for combining multi-modalities. The cross-entropy loss (CEL) may be included in three positions. "Dim" in M3Net2 represents the dimension of the feature with which will be used in the concatenation. The sub-paths in M3Net1 provide the intermediate estimated risks while sub-paths in M3Net2 provide high level features for concatenation.

The data distribution is illustrated in Table 1. The UPMC cohort is composed of patients with indeterminate pulmonary nodules, whose nodule size is between 6 and 30 mm.

## 2.2 Data Preprocessing and Nodule Detection

The data preprocessing follows the pipeline of Liao et al. [7] and adapts the open resource of https://github.com/lfz/DSB2017. In brief, as described in [7], the original CT is processed by (1) converting to Hounsfield Units, (2) binarizing by thresholding, (3) selecting the connected domain corresponding to the lungs, (4) segmenting left and right lungs, (5) computing the convex hull, (6) dilating two lung masks, (7) masking image, filling masked region with tissue luminance, converting to UINT8, (8) cropping image and clipping bone intensities. The detection of pulmonary nodules network and pre-trained model is also borrowed from Liao et al. [7]. Then, five regions with top confidence scores are selected for feature extraction If the number of detected nodule region is less than five, blank image sub-regions are padded to make the number nodule proposals as five. Validated by [7], the top five regions are enough to capture most nodules.

## 2.3 Multi-path Multi-modal Missing Network

We call the proposed model as multi-path multi-modal missing network (M3Net) as the model has multiple paths and is designed to handle missing modality of biomarkers and image. The framework is shown in Figure 2. The network path used for learning from CT image (the image-path) has two main parts. The Convolutional Neural Network (CNN) includes the nodule detection network and the feature extraction network from Liao et al. [7]. The output of the CNN is the five 128-dimension features respect to five nodules in this paper, which have the top five confidence scores. The AMIL represents Attention-based Multi-Instance Learning adapted from [21], which is a sub-net in our framework that feed by the five nodule-features. The input of the biomarker-path is a 10-dimension vector that contains the nine available biomarkers as Figure 1 and a higher-level factor Mayo risk. Two dense layers are applied to extract feature from biomarkers. The outputs from image-path and biomarker-path are concentrated with two other factors, blood and Mayo risk, which have been empirically validated as helpful to estimating cancer risk [3] [4]. The combined-path is composed of two dense layers and its output is the final cancer risk prediction. The reported performance is based on the final prediction of combined-path. The cross-entropy loss (CEL) is deployed in three positions in the pipeline for image-path, biomarker path and combined path, respectively. The image-path and biomarker-path can be trained and learned independently, as the motivation is to take advantage of the subjects with missing modality. There are three situations of the framework: (1) subject with image only: the M3Net will only include the Img-CEL and only the image-path will be trained; (2) subject with biomarkers only: the M3Net will only include the Bio-CEL and only the biomarker-path will be trained; (3) subject with both image and biomarkers: the three CELs will be included and all learnable parameters will be trained.

We provide two versions of M3Net, as shown in Figure 2, termed as M3Net1 and M3Net2. In M3Net1, intermediate estimated risks (i.e., $p_1, p_2$) which are supervised by CEL from image-path and biomarker-path have feed to the combine-path. Instead of intermediate estimated risk, the feature vector has been feed to the combine-path in M3Net2.

## 3. EXPERIMENTS AND RESULTS

### 3.1 Experimental Settings

Two experimental validation settings are applied in this paper: (1) cross-validation in VDD, and (2) external-validation in UPMC. The VDD cohort has been randomly split into five folds. In the cross-validation, each fold data has been held out from training as the test set, and remaining four folds are split as 3:1 for training and validation sets. In external validation, UPMC cohort is the held-out test set, four folds in VDD is the training set, and one-fold is the validation set. If the biomarkers acquisition date is missing, it would be matched with the last CT scan date by default.

Our experiments are implemented in Python 3.7 with PyTorch 1.5 using the GTX Titan X. The max-training-epoch is set to 100. The initial learning rate is 0.01 and is multiplied by 0.2 at the $40^{th}$, $60^{th}$, $80^{th}$ epochs. The optimizer used in the training is stochastic gradient descent (SGD).

### 3.2 Experimental Results

Table 2. The AUC of test set in cross-validation VDD and external-validation UPMC

| Methods | VDD | UPMC1 (mean ± std) | UPMC2 (CI) |
|---|---|---|---|
| Mayo Model† | 0.682* | 0.858* | 0.858 (0.776-0.926)* |
| Liao et al. pretrain model† | 0.663* | 0.810* | 0.810 (0.717-0.897)* |
| Image only | 0.712 ± 0.051 * | 0.863 ± 0.004 * | 0.863 (0.784-0.935)* |
| Biomarkers only | 0.757 ± 0.059 * | 0.826 ± 0.040 * | 0.852 (0.770-0.919)* |
| Learning without imperfect data | 0.793 ± 0.027 * | 0.886 ± 0.007 * | 0.889 (0.815-0.950)* |
| M3Net1 (ours) | 0.816 ± 0.048 | **0.913 ± 0.005** | **0.917 (0.857-0.966)** |
| M3Net2 (Dim=5) (ours) | **0.848 ± 0.052** | 0.910 ± 0.011 | 0.916 (0.851-0.968) |

The models with † do not need re-training in this paper.

The * represents that our method M3Net1 significantly improve the performance compared with other methods ($p < 0.05$, bootstrap two-tailed test).

The UPMC1 represents that the reported result on UPMC cohort is the average and standard deviation of the AUC values of five-fold models from training. The UPMC2 represents the reported result on UPMC cohort is the AUC value and its 95% confidential interval (CI) where the predicted cancer probability for each subject is the average predicted cancer probabilities of five-fold models.

The test-set results of cross-validation and external-validation are shown in Table 2. The results are reported on the subjects with both biomarkers and CT in Table 1. The compared benchmarks including (1) the methods Mayo Model [3] and Liao et al. [7] pre-trained model, which does not need training in this paper, (2) single modality prediction of image only and biomarkers only. The single modality prediction has the same training procedure and training data of the proposed method, but the evaluation is based on the $p_1$ and $p_2$ of M3Net1 in Figure 2, respectively. (3) Learning while excluding the subjects that with missing modality which is termed as *Learning without imperfect data* in Table 2. This baseline has the same network structure as M3Net1, the only difference being that the training data only includes those subjects with both CT image and biomarkers.

Our method significantly improves the compared benchmarks ($p < 0.05$, bootstrap two-tailed test). The computing of p-value and 95% confidential interval (CI) is adapted from: https://github.com/mateuszbuda/ml-stat-util.

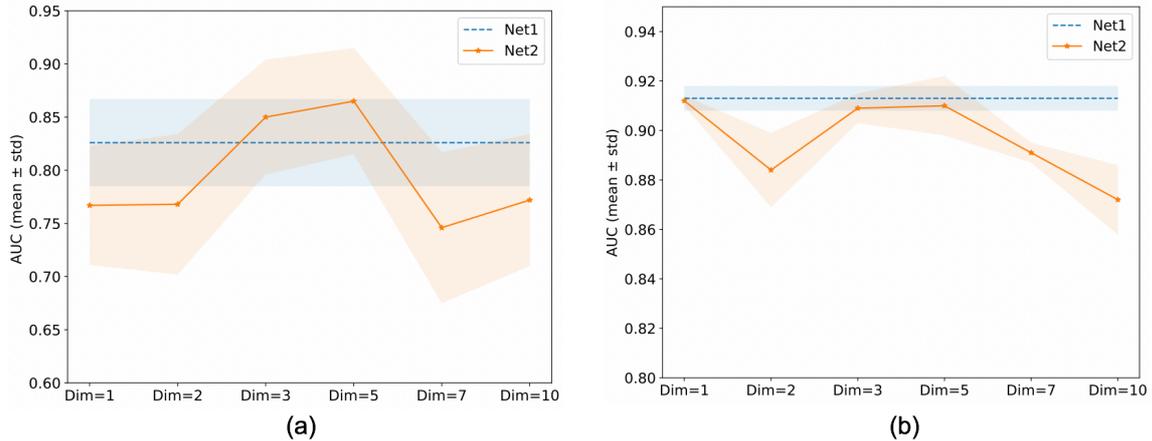

**Fig 3.** The comparison of M3Net1 and M3Net2 in the validation set and test set in the external validation setting, where AUC (mean ± std) of five folds is shown: (a) The performance on validation set (VDD). (b) The performance on test set (UPMC).

Figure 3 shows the comparison of M3Net1 and M3Net2 in the validation set and test set in the external-validation setting. In M3Net2, we compare the settings of different "Dim" parameters. In general, the performance of M3Net1 is comparable with the best performance of M3Net2 under our settings and datasets. The performance decreases when "Dim" arises, which probably results from that the network being unable to learn effective large-dimension feature given the limited data size. $p_1$

## 4. DISCUSSION

In this paper, we propose a new framework M3Net to estimate lung cancer risk by integrating multiply modalities with missing data. The proposed network shows superior performance compared with single modality predicting and learning without imperfect data, which indicates the combining multi-modalities and utilizing more subjects even with missing data can increase the performance of lung cancer diagnosis. We compare two versions of our model with their performance on validation set and test set. A limitation of this paper is the data size is not as large as general classification tasks in computer vision field, therefore, the generalization across different patient cohorts requires care.

## 5. ACKNOWLEDGEMENT

This research was supported by NSF CAREER 1452485 and R01 EB017230. This study was supported in part by a UO1 CA196405 to Massion. This study was in part using the resources of the Advanced Computing Center for Research and Education (ACCRE) at Vanderbilt University, Nashville, TN. This project was supported in part by the National Center for Research Resources, Grant UL1 RR024975-01, and is now at the National Center for Advancing Translational Sciences, Grant 2 UL1 TR000445-06. We gratefully acknowledge the support of NVIDIA Corporation with the donation of the Titan X Pascal GPU used for this research. The de-identified imaging dataset(s) used for the analysis described were obtained from ImageVU, a research resource supported by the VICTR CTSA award (ULTR000445 from NCATS/NIH), Vanderbilt University Medical Center institutional funding and Patient-Centered Outcomes Research Institute (PCORI; contract CDRN-1306-04869). This study was funded in part by the Martineau Innovation Fund Grant through the Vanderbilt-Ingram Cancer Center Thoracic Working Group and NCI Early Detection Research Network 2U01CA152662.

## 6. REFERNCES